\documentclass[preprint]{revtex4}

\usepackage{amsfonts}
\usepackage{amsmath}
\usepackage{amssymb}
\usepackage{graphicx}

\begin{document}
\title{Time-reversal of a wave packet emitted by an internal source: the perfect inverse filter}

\author{Hern\'{a}n L. Calvo}
\affiliation{Facultad de Matem\'{a}tica, Astronom\'{i}a y F\'{i}sica, Universidad Nacional de C\'{o}rdoba, Ciudad Universitaria, 5000 C\'{o}rdoba, Argentina.}

\author{Horacio M. Pastawski}
\affiliation{Facultad de Matem\'{a}tica, Astronom\'{i}a y F\'{i}sica, Universidad Nacional de C\'{o}rdoba, Ciudad Universitaria, 5000 C\'{o}rdoba, Argentina.}

\begin{abstract}
The time-reversal mirror (TRM) of acoustic waves is used as a focusing technique that is robust against disorder and chaos. The propagation of an ultrasonic pulse from the focal 
point is detected and recorded by a surrounding array of transducers that enclose the cavity region. The registered signals are then played-back in the reverse temporal sequence 
and an imperfect Loschmidt echo is obtained. A similar technique developed in the quantum domain, the perfect inverse filter (PIF), ensures an exact reversion. Applied to the 
acoustical case, it allows an improvement in the design of a minimally invasive inverse filter. Although that version of PIF accounts for external source condition, now we 
propose a generalization where the initial excitation is created inside the cavity. We present the theoretical principles for this form of PIF and perform a numerical test in a 
multichannel system that improves the focalization quality compared to the TRM version.
\end{abstract}

\maketitle

\section{Introduction}

In acoustic experiments, the time-reversal mirror (TRM) technique has been developed by Mathias Fink and collab. \cite{Fink} in order to focus ultrasonic waves through 
heterogeneous media. The remarkable robustness of this technique against inhomogeneities allows novel applications in medical physics \cite{Montaldo}, oceanography \cite{Edelman}
and telecommunications \cite{telecom}. The acoustic signals diverging from a localized source point or scatterer are recorded by an array of transducers until all excitations 
have decayed. Their registries are then played-back following the inverted time sequence. Notably, with this simple prescription, the initial excitation is recovered in the 
source region up to a good degree of fidelity. This constitutes an alternative form of achieving the Loschmidt Echo \cite{Jalabert}, i.e. the time reversal of an excitation that 
has spread in a complex medium. In its closed version, called time-reversal cavity, the array of transducers $B$ encloses the source region separating the space in two portions: 
the closed cavity $A$, where the focusing takes place; and the outer region, towards which all the excitations finally escapes. In order to ensure the perfect refocusing and 
quantify the stability of the TRM, we have introduced a formal prescription for quantum systems denoted as the perfect inverse filter (PIF) \cite{Pastawski}. This achieves exact 
focusing by injecting a wave function that has to be corrected to account for the time reversed signal and the system response. The injection prescription was derived as a 
solution to the Schr\"{o}dinger equation with boundary source conditions that arises when a wave packet arrives at the cavity from the outside. As the wave packet enters at the 
cavity, it interferes with its reflected component and finally spreads out in space and time. The reversed counterpart is obtained via the injection process and a burst of wave 
probability is emitted from $A$. This is the quantum version of the \textit{sound bazooka}, suggested by M. Fink and collaborators \cite{Silver}. Furthermore, we implemented 
this prescription in a classical chain of harmonic oscillators \cite{Calvo}. In that case, a low frequency mode (e.g. a heavy mass) was coupled to a semi-infinite chain of 
lighter masses leading to broad continuum spectrum of excitations that acts as an "environment". The energy stored in the surface oscillator decays almost exponentially and its 
propagation was registered by a single transducer inside the environment. In this case where all the normal modes participate in the local excitation, the PIF correction becomes 
effective and both the dynamical build-up and the final contrast, spatial distribution of the recovered initial condition, are substantially improved.

In addition to the TRM, Fink and collab. proposed the spatio-temporal inverse filter (STIF) for acoustic waves as a focusing method that optimizes spatial contrast 
\cite{STIF}. It requires the inversion of the Fourier transform of the direct wavefield propagator. This matrix is build from the temporal response at the transducers to a Dirac
delta function in each control point within the focal region. Unlike the TRM, the STIF injection acts as a highly invasive technique because it is necessary to measure the full
propagator whereas the former only need to compute the row related to the focal point. Recently, an alternative proposal by \textit{Vignon et al.} \cite{Vignon}, denoted 
minimally invasive STIF (miSTIF), achieves the same focusing quality than STIF. The idea behind this procedure is that all the information in the wavefield propagator can be 
deduced from the backscattering signals of the transducer array at the boundaries. Hence the technique is no more invasive than TRM. In this context, miSTIF and PIF result 
similar because they correct the TRM injection function using the propagator in the reduced basis of the transducers.

In this work, we explore the PIF procedure for the case in which the source is placed inside the cavity. In Sec. II the general PIF formula is deduced from the properties of the 
system Green's function. The obtained result is then applied in a semi-infinite chain with a single transducer and also to a general multichannel system. For this last case, a 
numerical simulation is performed yieding a favourable test of the most difficult task: the time reversal of an initially localized excitation. In Sec. III the PIF is reobtained 
from the optical theorem and compared with the miSTIF technique.

\section{Time-reversal with internal source}

We look for the situation in which at the initial time \textit{all} excitations are located \textit{inside} the cavity. If we denote by $\psi_{x}(t_{0})$ the coefficient of the 
initial condition at site $x$, then we have
\begin{equation}
\vec{\psi}(t_{0})=\vec{\psi}_{A}(t_{0})\equiv\sum_{x\in A}\psi_{x}(t_{0})\left\vert x\right\rangle .
\end{equation}
During the propagation from $t_{0}=0$ until a registration time $t_{R}$, the wave function is detected and recorded as crosses the boundaries $B$
\begin{equation}
\vec{\psi}_{B}(t)\equiv\sum_{x\in B}\psi_{x}(t)\left\vert x\right\rangle=\mathrm{i}\hbar\mathbb{G}_{B,A}(t)\vec{\psi}_{A}(0),\text{ }0\leq t\leq t_{R},
\label{eq_for}
\end{equation}
where $\mathbb{G}_{B,A}(t)$ is a matrix arrangement whose element $G(x,t;x^{\prime},0)$ relates the response at site $x\in B$ to a Dirac delta function $\delta(t)$ in 
$x^{\prime}\in A$ following the Schr\"{o}dinger equation whose Hamiltonian is $\mathbb{H}$
\begin{equation}
\mathrm{i}\hbar\frac{\partial}{\partial t}G(x,t;x^{\prime},0)-\sum_{x^{\prime\prime}}H_{x,x^{\prime\prime}}G(x^{\prime\prime},t;x^{\prime},0)=\delta_{x,x^{\prime}}\delta(t).%
\end{equation}
The Fourier transform of this matrix gives the retarded Green's function
\begin{equation}
G(x,x^{\prime};\varepsilon)=\left[  (\varepsilon\mathbb{I}-\mathbb{H})^{-1}\right]  _{x,x^{\prime}}.
\label{eq_green}
\end{equation}
Assuming that there are no localized states, the registration time $t_{R}$ is taken long enough in order to ensure that all excitations have been left the cavity 
($\vec{\psi}_{A}(t_{R})\simeq0$).

As we have shown in Ref.\cite{Pastawski}, the time reversion for a wave packet arriving at the cavity from the outer region is performed through the injection
\begin{equation}
\vec{\chi}_{B}(\varepsilon)=\frac{1}{\mathrm{i}\hbar}\left[  \mathbb{G}_{B,B}(\varepsilon)\right]  ^{-1}\vec{\psi}_{B}^{\mathrm{rev}}(\varepsilon).
\end{equation}
Notice that in the above expression,
\begin{equation}
\vec{\psi}_{B}^{\mathrm{rev}}(\varepsilon)\equiv\int_{-\infty}^{\infty}\mathrm{d}t\exp(\mathrm{i}\varepsilon t/\hbar)\vec{\psi}_{B}^{\mathrm{rev}}(t),
\end{equation}
accounts for the time reversed \textit{complete evolution} of the wave packet, i.e. both the incoming and outgoing components are necessary in order to compute 
$\vec{\psi}_{B}^{\mathrm{rev}}(\varepsilon)$. However, in the present work we want to reverse the signal that is produced inside the cavity. Then, we should deal with the 
building up of the complete evolution from the knowledge of only a partial component $\vec{\psi}_{B}(t)$ at times $0\leq t\leq t_{R}$. By supposing that the injection function 
is known, the reversed propagation at the boundary should be
\begin{equation}
\vec{\psi}_{B}^{\mathrm{rev}}(t)=\vec{\psi}_{B}^{\ast}(t_{R}-t)=\mathrm{i}\hbar\int_{0}^{t}\mathbb{G}_{B,B}(t-t^{\prime})\vec{\chi}_{B}(t^{\prime})\mathrm{d}t^{\prime},\ \ 0\leq t\leq t_{R}.
\end{equation}
Since the PIF prescription achieves for exact reversion inside the cavity, the evolution at subsequent times can be assumed as the wave packet that starts at the focalization 
time with $\vec{\psi}_{A}(t_{R})=\vec{\psi}_{A}^{\ast}(0)$. Thus, for subsequent times we have
\begin{equation}
\vec{\psi}_{B}^{\mathrm{rev}}(t)=\mathrm{i}\hbar\mathbb{G}_{B,A}(t-t_{R})\vec{\psi}_{A}^{\ast}(0),\ t>t_{R}. \label{eq_subs}
\end{equation}
A comparison between Eqs.(\ref{eq_for}) and (\ref{eq_subs}) shows that for the most simple case in which all coefficients of $\vec{\psi}_{A}(0)$ are real, both evolutions become
identical. The same evolution but sign changed should be obtained when $\vec{\psi}_{A}(0)$ is purely imaginary.

In a more general situation where the coefficients present relative phases we start with the Fourier transform
\begin{subequations}
\begin{align}
\vec{\psi}_{B}(\varepsilon)  &  =\mathrm{i}\hbar\mathbb{G}_{B,A}(\varepsilon)\vec{\psi}_{A}(0),\\
\vec{\psi}_{B}^{\mathrm{rev}}(\varepsilon)  &  =\int_{0}^{t_{R}}\vec{\psi}_{B}^{\ast}(t_{R}-t)e^{\mathrm{i}\varepsilon t/\hbar}\mathrm{d}t+\mathrm{i}\hbar\int_{t_{R}}^{\infty}\mathbb{G}_{B,A}(t-t_{R})\vec{\psi}_{A}^{\ast}(0)e^{\mathrm{i}\varepsilon t/\hbar}\mathrm{d}t\\
&  =e^{\mathrm{i}\varepsilon t_{R}/\hbar}\left[  \vec{\psi}_{B}^{\ast}(\varepsilon)+\mathrm{i}\hbar\mathbb{G}_{B,A}(\varepsilon)\vec{\psi}_{A}^{\ast}(0)\right]  , \label{eq_rev}
\end{align}
\end{subequations}
where the second term within the brackets can be interpreted as the unknown evolution for subsequent times. At this point, we account for the hopping matrices $\mathbb{V}_{B,A^{\prime}}$
and $\mathbb{V}_{A^{\prime},B}$ ($A^{\prime}\subseteq A$) connecting the cavity with the outer region. The $\varepsilon$ argument in $\mathbb{G}_{B,A}(\varepsilon)$ is neglected 
for a compact notation henceforth. Hence, it is possible to rewrite $\mathbb{G}_{B,A}$ in terms of the Dyson equation
\begin{equation}
\mathbb{G}_{B,A}=\mathbb{G}_{B,A}^{(0)}+\mathbb{G}_{B,B}\mathbb{V}_{B,A^{\prime}}\mathbb{G}_{A^{\prime},A}^{(0)},
\end{equation}
and its transpose
\begin{equation}
\mathbb{G}_{A,B}\equiv\left[  \mathbb{G}_{B,A}\right]  ^{T}=\mathbb{G}_{A,B}^{(0)}+\mathbb{G}_{A,A^{\prime}}^{(0)}\mathbb{V}_{A^{\prime},B}\mathbb{G}_{B,B},
\end{equation}
where $\mathbb{G}^{(0)}$ follows Eq.(\ref{eq_green}) for the unperturbed Hamiltonian with $\mathbb{V}_{B,A^{\prime}}=\mathbb{V}_{A^{\prime},B}=0$. For the considered Green's 
functions we have $\mathbb{G}_{B,A}^{(0)}=0$ and since the subsystem $A$ is closed and non absorbing, $\mathbb{V}_{B,A^{\prime}}\mathbb{G}_{A^{\prime},A}^{(0)}$ is a real matrix.
Therefore, we obtain for the advanced Green's function
\begin{equation}
\left[  \mathbb{G}_{B,A}\right]  ^{\dag}\equiv\mathbb{G}_{A,B}^{\ast}=\mathbb{G}_{A,A^{\prime}}^{(0)}\mathbb{V}_{A^{\prime},B}\mathbb{G}_{B,B}^{\ast}=\mathbb{G}_{A,B}\left[  \mathbb{G}_{B,B}\right]  ^{-1}\left[ \mathbb{G}_{B,B}\right]  ^{\dag},
\end{equation}
and
\begin{subequations}
\begin{align}
\mathrm{i}\hbar\mathbb{G}_{B,A}\vec{\psi}_{A}^{\ast}(0)  &  =\left[ -\mathrm{i}\hbar\vec{\psi}_{A}^{T}(0)\mathbb{G}_{A,B}^{\ast}\right]  ^{\dag}=\left[  -\mathrm{i}\hbar\vec{\psi}_{A}^{T}(0)\mathbb{G}_{A,B}\left[ \mathbb{G}_{B,B}\right]  ^{-1}\left[  \mathbb{G}_{B,B}\right]  ^{\dag}\right] ^{\dag}\\
&  =-\mathbb{G}_{B,B}\left[  \left[  \mathbb{G}_{B,B}\right]  ^{-1}\right] ^{\dag}\vec{\psi}_{B}^{\ast}(\varepsilon).
\end{align}
\end{subequations}
Finally, we use this last expression and rewrite Eq.(\ref{eq_rev}) as
\begin{equation}
\vec{\psi}_{B}^{\mathrm{rev}}(\varepsilon)=e^{\mathrm{i}\varepsilon t_{R}/\hbar}\left[  \mathbb{I}-\mathbb{G}_{B,B}\left[  \left[  \mathbb{G}_{B,B}\right]  ^{-1}\right]  ^{\dag}\right]  \vec{\psi}_{B}^{\ast}(\varepsilon).
\end{equation}
Now we have the complete $\vec{\psi}_{B}^{\mathrm{rev}}(\varepsilon)$, the injection function according to the PIF formula is given by
\begin{equation}
\vec{\chi}_{B}(\varepsilon)=\frac{1}{\mathrm{i}\hbar}\left[  \mathbb{G}_{B,B}\right]  ^{-1}\vec{\psi}_{B}^{\mathrm{rev}}(\varepsilon)=\frac{2}{\hbar}\exp(\mathrm{i}\varepsilon t_{R}/\hbar)\operatorname{Im}\left(  \left[ \mathbb{G}_{B,B}\right]  ^{-1}\right)  \vec{\psi}_{B}^{\ast}(\varepsilon).\label{eq_PIF}
\end{equation}
This is the PIF prescription for the case where the initial state is an excitation inside the cavity. As we shall see, the imaginary component of the inverse in 
$\mathbb{G}_{B,B}$ is closely related with the group velocity of the scattered waves in the outer region. Therefore, contrary with the external source condition presented in 
Ref.\cite{Pastawski}, the correction imposed by the PIF in internal source condition do not depends on the structure of the cavity.

\subsection{Application examples}

As a first attempt to evaluate Eq.(\ref{eq_PIF}) we propose a one-dimensional discrete system whose Hamiltonian writes
\begin{equation}
H=\sum_{x\geq0}E_{x}\left\vert x\right\rangle \left\langle x\right\vert+\sum_{x\geq0}(V_{x,x+1}\left\vert x+1\right\rangle \left\langle x\right\vert+h.c.),
\end{equation}
where $E_{x}$ denotes the site energy at $x\geq0$ and $V_{x,x+1}$ is the hopping term between sites $x$ and $x+1$. In this system, the cavity can be delimited by only one 
transducer at site $x_{S}$. The Green's function is obtained through the continued fraction technique as
\begin{equation}
G(x_{S},x_{S};\varepsilon)=\frac{1}{\varepsilon-E_{x_{S}}-\Sigma_{\mathrm{in}}(\varepsilon)-\Sigma_{\mathrm{out}}(\varepsilon)},
\end{equation}
where $\Sigma_{\mathrm{in}}(\varepsilon)$ and $\Sigma_{\mathrm{out}}(\varepsilon)$ are the self-energy corrections to the site energy $E_{x_{S}}$
due to the presence of the remaining sites inside and outside the cavity respectively \cite{Medina}.

Decimation on the cavity gives the $\Sigma_{\mathrm{in}}(\varepsilon)$ contribution as a continued fraction whose parameters are the hoppings and the site energies and, in 
absence of magnetic fields and absorption processes, they are all real. Therefore,
\begin{equation}
\Sigma_{\mathrm{in}}(\varepsilon)=V_{x_{S},x_{S-1}}\frac{1}{\varepsilon-E_{x_{S-1}}-V_{x_{S-1},x_{S-2}}\dfrac{1}{\varepsilon-E_{x_{S-2}}-\cdots V_{x_{1},x_{0}}\dfrac{1}{\varepsilon-E_{0}}V_{x_{0},x_{1}}}V_{x_{S-2},x_{S-1}}}V_{x_{S-1},x_{S}},
\end{equation}
is also a real function.

In the thermodynamic limit where the number of sites outside increases indefinitely, the homogeneous outer region gives the self energy as a complex number:
\begin{equation}
\Sigma_{\mathrm{out}}(\varepsilon)=\Delta_{\mathrm{out}}(\varepsilon)-\mathrm{i}\Gamma_{\mathrm{out}}(\varepsilon),
\end{equation}
with
\begin{equation}
\Delta_{\mathrm{out}}(\varepsilon)\equiv\Delta(\varepsilon-E_{\mathrm{O}})=\left\{
\begin{array}
[c]{cc}
\frac{\varepsilon-E_{\mathrm{O}}}{2}+\sqrt{\left(  \frac{\varepsilon-E_{\mathrm{O}}}{2}\right)  ^{2}-V^{2}} & \varepsilon-E_{\mathrm{O}}<-2\left\vert V\right\vert \\
\frac{\varepsilon-E_{\mathrm{O}}}{2} & \left\vert \varepsilon-E_{\mathrm{O}}\right\vert \leq2\left\vert V\right\vert \\
\frac{\varepsilon-E_{\mathrm{O}}}{2}-\sqrt{\left(  \frac{\varepsilon-E_{\mathrm{O}}}{2}\right)  ^{2}-V^{2}} & \varepsilon-E_{\mathrm{O}}>2\left\vert V\right\vert
\end{array}
\right.,
\label{eq_delta}
\end{equation}
and
\begin{equation}
\Gamma_{\mathrm{out}}(\varepsilon)\equiv\Gamma(\varepsilon-E_{\mathrm{O}})=\left\{
\begin{array}
[c]{cc}
\sqrt{V^{2}-\left(  \frac{\varepsilon-E_{\mathrm{O}}}{2}\right)  ^{2}} &
\left\vert \varepsilon-E_{\mathrm{O}}\right\vert \leq2\left\vert V\right\vert\\
0 & \left\vert \varepsilon-E_{\mathrm{O}}\right\vert >2\left\vert V\right\vert
\end{array}
\right.,
\label{eq_gamma}%
\end{equation}
whose parameters $V$ and $E_{\mathrm{O}}$ are the hopping and the site energy in the outer region respectively. The relationship between $\Gamma_{\mathrm{out}}(\varepsilon)$ and 
the group velocity can be found through the dispersion relation for the asymptotic waves:
\begin{subequations}
\begin{align}
\varepsilon_{k}  &  =E_{\mathrm{O}}-2V\cos(ka),\label{eq_disp}\\
v_{g}  &  =\frac{1}{\hbar}\frac{\mathrm{d}\varepsilon_{k}}{\mathrm{d}k}=\frac{2V}{\hbar}a\sin(ka)=\frac{2a}{\hbar}\sqrt{V^{2}-\left( \frac{\varepsilon_{k}-E_{\mathrm{O}}}{2}\right) ^{2}}\\
&  =\frac{2a}{\hbar}\Gamma_{\mathrm{out}}(\varepsilon_{k}),
\end{align}
\end{subequations}
with $a$ the distance between first neighbor sites. Remember that choosing $E_{\mathrm{O}}=2V$ so that the differential form of the Schr\"{o}dinger equation for a particle with 
mass $m$ is obtained as a limit $\hbar^{2}/2Va^{2}\longrightarrow m$ when $V\rightarrow\infty$ and $a^{2}\rightarrow0$. Since the only imaginary component from 
$\Sigma_{\mathrm{out}}(\varepsilon)$ is required, Eq.(\ref{eq_PIF}) writes
\begin{equation}
\chi_{\mathrm{PIF}}(x_{S},\varepsilon)\equiv\chi_{B}(x_{S},\varepsilon)=\frac{2}{\hbar}e^{\mathrm{i}\varepsilon t_{R}/\hbar}\Gamma_{\mathrm{out}}(\varepsilon)\psi^{\ast}(x_{S},\varepsilon).
\end{equation}
The complex exponential serves to define the origin of time, and hence we can neglect it. Using the definition of $\Gamma(\varepsilon)$ given above we get
\begin{subequations}
\begin{align}
\chi_{\mathrm{PIF}}(x_{S},\varepsilon)  &  =\sqrt{1-\left(  \frac
{\varepsilon-2V}{2V}\right)  ^{2}}\tfrac{2V}{\hbar}\psi^{\ast}(x_{S}%
,\varepsilon)\\
&  =\underset{\mathrm{perfect\ inverse\ filter}%
}{\underbrace{\sqrt{1-\left(  \frac{\varepsilon-2V}{2V}\right)  ^{2}}}}%
\chi_{\mathrm{TRM}}(x_{S},\varepsilon),
\end{align}
\end{subequations}
i.e., the time reversal mirror prescription of injecting an amplitude proportional to the detected wave function must be corrected by a filter that enables the exact reversal. 
When the initial state is a local excitation composed by a few sites inside the cavity, the Fourier transform of the detected signal $\psi(x_{S},\varepsilon)$ will cover the 
whole energy band between $-2V$ and $2V$. The factor in the PIF procedure shows that the correction becomes effective near the band edges where $\Gamma_{\mathrm{out}}\sim0$. 
However, since in simple homogeneous cases the filter is just proportional to the group velocity, it is clear that in the acoustical case 
$\chi_{\mathrm{PIF}}(x_{S},\varepsilon)=\chi_{\mathrm{TRM}}(x_{S},\varepsilon).$

A more general case where Eq.(\ref{eq_PIF}) can be tested is the multichannel system schematically shown in Fig.\ref{fig_scheme}. Here, a two dimensional cavity is surrounded by 
$N_{L}$ transducers at the left and $N_{R}$ at right. The total Hamiltonian is written as
\begin{equation}
\mathbb{H=}\left(
\begin{array}
[c]{ccc}%
\mathbb{H}_{L} & \mathbb{V}_{L}(\varepsilon) & 0\\
\mathbb{V}_{L}^{T}(\varepsilon) & \mathbb{H}_{C} & \mathbb{V}_{R}%
(\varepsilon)\\
0 & \mathbb{V}_{R}^{T}(\varepsilon) & \mathbb{H}_{R}%
\end{array}
\right)  ,
\end{equation}
where $\mathbb{H}_{C}$ is the Hamiltonian inside the cavity and $\mathbb{H}_{L}$ ($\mathbb{H}_{R}$) is the Hamiltonian on the left \ (right) side. $\mathbb{V}_{L}(\varepsilon)$ 
and $\mathbb{V}_{R}(\varepsilon)$ are real matrices containing the hopping terms between the boundaries and the cavity. The unitary transform
\begin{equation}
\mathbb{U}=\left(
\begin{array}
[c]{ccc}%
\mathbb{U}_{_{L}} & 0 & 0\\
0 & \mathbb{I}_{C} & 0\\
0 & 0 & \mathbb{U}_{R}%
\end{array}
\right) ,
\end{equation}
with $\mathbb{I}_{C}$ the identity matrix in the cavity, separates the channels into $N_{L}+N_{R}$ independent one-dimensional homogeneous chains. The energies $E_{k}^{\alpha}$ 
follow the dispersion relation Eq.(\ref{eq_disp}) for the discrete case:
\begin{equation}
E_{k}^{\alpha}=E_{\mathrm{O}}-2V\cos\left(  \frac{k\pi}{N_{\alpha}+1}\right)
,\text{ }k=1,\ldots,N_{\alpha}.
\end{equation}
and
\begin{equation}
[\mathbb{U}_{\alpha}]_{nk}=u_{nk}^{\alpha}=u_{kn}^{\alpha}=\sqrt{\frac{2}{N_{\alpha}+1}}\sin\left(
\frac{nk\pi}{N_{\alpha}+1}\right) .
\end{equation}
By performing a decimation process in each one of these chains, the effective site energy at the boundaries is
\begin{equation}
\left[  \mathbb{U}_{\alpha}^{\dag}\mathbb{H}_{\alpha}\mathbb{U}_{\alpha
}\right]  _{k,k^{\prime}}\mathbb{=}\left[  E_{k}^{\alpha}-\Sigma
(\varepsilon-E_{k}^{\alpha})\right]  \mathbb{\delta}_{k,k^{\prime}},
\label{eq_diag}%
\end{equation}
where $\alpha=L,R\ $and$\ k,k^{\prime}\in B$ denotes the eigenstate index in the transducers subsystem. We use again the decimation procedure on the cavity sites in order to 
obtain the total Hamiltonian in the reduced basis of the boundaries $\mathbb{H}_{B,B}(\varepsilon)$. Since the Green's function is defined as:
\begin{equation}
\mathbb{G}_{B,B}\equiv\left[  \left(  \varepsilon\mathbb{I}-\mathbb{H}\right)
^{-1}\right]  _{B,B}=\mathbb{U}\left(  \varepsilon\mathbb{I}_{B,B}%
-\mathbb{H}_{B,B}(\varepsilon)\right)  ^{-1}\mathbb{U}^{\dag},
\end{equation}
we have,
\begin{subequations}
\begin{align}
\left[  \mathbb{G}_{B,B}\right]  ^{-1}  &  =\mathbb{U}\left[  \varepsilon
\mathbb{I}_{B,B}-\mathbb{H}_{B,B}(\varepsilon)\right]  \mathbb{U}^{\dag},\\
\operatorname{Im}\left[  \mathbb{G}_{B,B}\right]  ^{-1}  &  =-\mathbb{U}%
\operatorname{Im}\left[  \mathbb{H}_{B,B}(\varepsilon)\right]  \mathbb{U}%
^{\dag}.
\end{align}
\end{subequations}
The only imaginary components in $\mathbb{H}_{B,B}(\varepsilon)$ come from the diagonal part and, as indicated in Eq.(\ref{eq_diag}),
\begin{equation}
\operatorname{Im}\left[  \mathbb{G}_{B,B}\right]  _{\left(  n,l\right)
\in\alpha}^{-1}=\sum_{k}u_{nk}^{\alpha}\Gamma(\varepsilon-E_{k}^{\alpha
})u_{kl}^{\alpha}, \label{eq_multichannel}%
\end{equation}
where $\left( i,j\right) \in\alpha$ means that both $n$ and $l$ are taken in the same side of the system: mixing terms between $L$ and $R$ sides of the above matrix should be 
zero. Thus, the injection function is obtained replacing Eq.(\ref{eq_multichannel}) in Eq.(\ref{eq_PIF}) as,
\begin{equation}
\left[  \vec{\chi}_{B,\alpha}(\varepsilon)\right]  _{n}=\frac{2}{\hbar
}e^{\mathrm{i}\varepsilon t_{R}/\hbar}\sum_{k,l=1}^{N_{\alpha}}u_{nk}^{\alpha
}\Gamma(\varepsilon-E_{k}^{\alpha})u_{kl}^{\alpha}\left[  \vec{\psi}%
_{B,\alpha}^{\ast}(\varepsilon)\right]  _{l}.
\end{equation}

As we have seen previously, the PIF correction for the internal source case is given by the group velocity at the boundaries. Notice that the $\Gamma(\varepsilon)$ factors are 
centered on the eigen-energies $E_{k}^{\alpha}$ whose width is given by the hoppings in the outer region. As we show in the numerical simulation, depending on the particular 
choise of these parameters the correction would be non trivial.

\subsection{Numerical simulation}

\begin{figure}
\begin{center}
\includegraphics{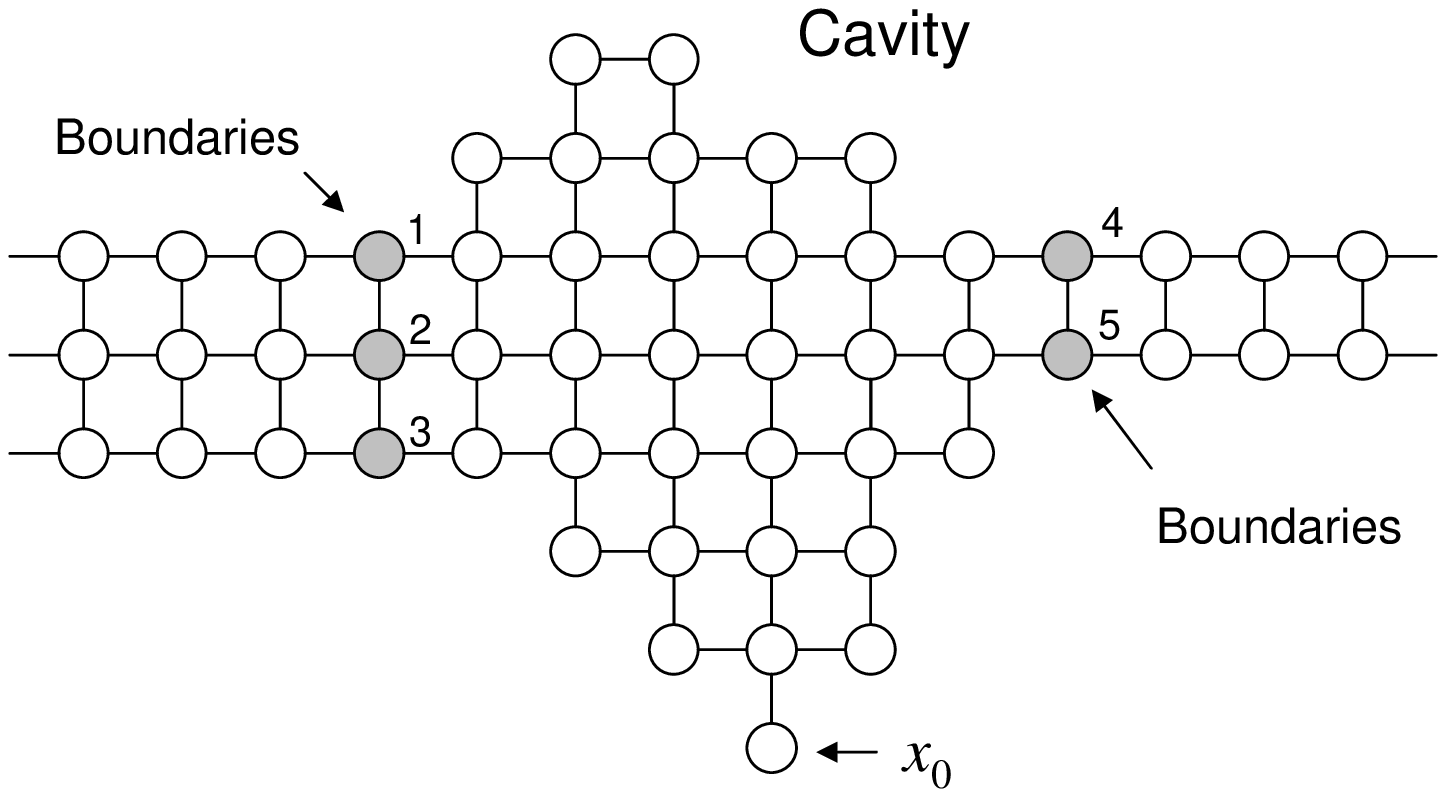}
\caption{Scheme of the multichannel system. White circles: cavity and channels sites. Gray circles: detection/injection sites (transducers). 
Lines connecting circles represent the hoppings terms.}
\label{fig_scheme}
\end{center}
\end{figure}

In the multichannel system depicted in Fig.\ref{fig_scheme} we want to reverse the dynamic that follows an initial condition of local excitation in the site labeled as $x_{0}$. We 
perform a numerical simulation for the time domain using the Trotter-Suzuki algorithm \cite{De Raedt}. In such case we have $N_{L}=3$ transducers for the left and $N_{R}=2$ for 
the right and the propagation is favored through the vertical hoppings (according to Fig.\ref{fig_scheme}) of the channels whose values are twice the horizontal ones.

\begin{figure}
\begin{center}
\includegraphics{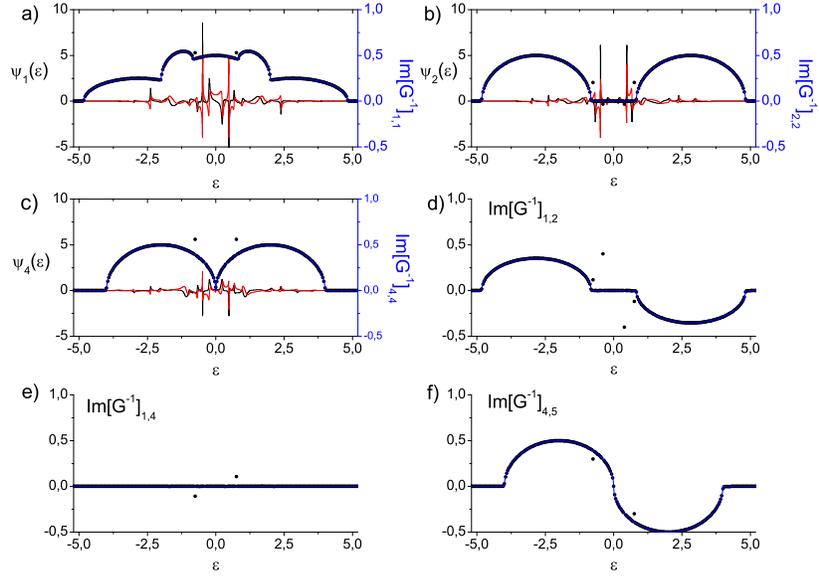}
\caption{(Color Online) Representative elements of the PIF correcion: numerical in black points and analytical in blue line. Real and imaginary components of
$\psi(x_{S},\varepsilon)$ are shown for comparison reasons.}
\label{fig_PIFcorrection}
\end{center}
\end{figure}

\begin{figure}
\begin{center}
\includegraphics{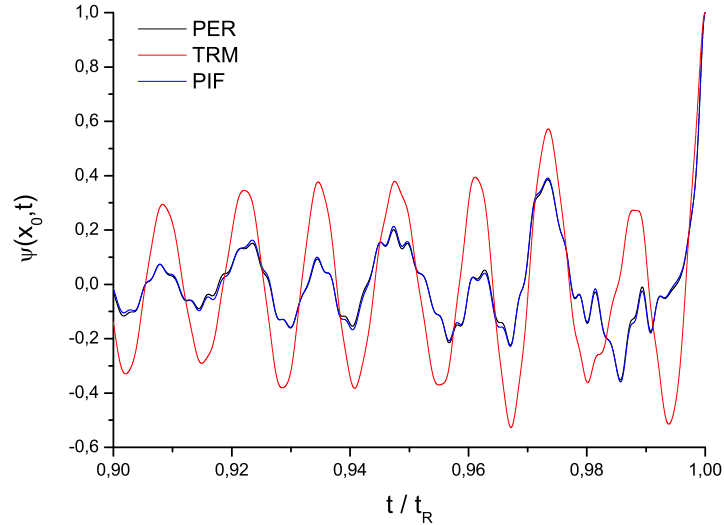}
\caption{(Color online) Recovering of the initial condition in the site $x_{0}$. Black line: ideal reversed evolution. Red line: normalized TRM. Blue
line: PIF.}
\label{fig_recovering}
\end{center}
\end{figure}

\begin{figure}
\begin{center}
\includegraphics{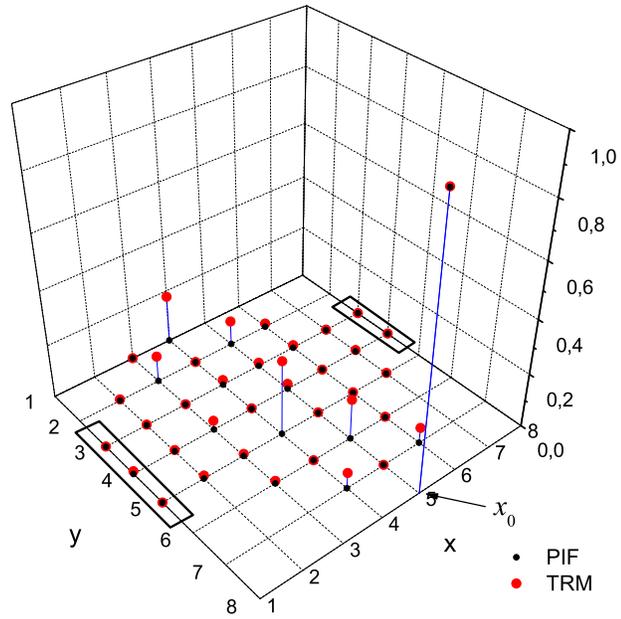}
\caption{(Color online) Normalized spatial distribution for the recovered initial condition in the cavity. The location of the transducers is indicated
by rectangles.}
\label{fig_spatial}
\end{center}
\end{figure}

The time-dependent response functions $G(x_{S^{\prime}},t;x_{S},0)$ between the transducers at the boundaries are then calculated and a numerical Fourier Transform is performed 
in order to obtain the $5\times5$ matrix $\mathbb{G}_{B,B}(\varepsilon)$. A comparison between analytical and numerical evaluation of Eq.(\ref{eq_multichannel}) is shown in 
Fig.\ref{fig_PIFcorrection} for some representative elements. They show how the group velocities in each channel are mixed. Figures \ref{fig_PIFcorrection}a, 
\ref{fig_PIFcorrection}b and \ref{fig_PIFcorrection}c show the diagonal elements and the Fourier transform of the detected signals for transducers 1,2 and 4 respectively. For
transducers 2 and 4, the correction becomes effective in the central region of the spectrum where the detected signals are appreciable. As we observed from 
Eq.(\ref{eq_multichannel}), Fig.\ref{fig_PIFcorrection}e shows that no mixing exist between transducers that belongs to different sides of the cavity. We can observe a little 
portion of points from the numerical results that does not fit with the analytical curves. These errors are related with the approximation done by the Trotter-Suzuki and because a 
finite registration time was taken. Once we have the corrected injection, we reverse the dynamic and compare it with the TRM case. Although TRM\ achieves a reasonable quality in 
the spatial focalization, differences appear both for the time evolution at the focal site (see Fig.\ref{fig_recovering}) and the spatial distribution of the recovered signal at 
the focalization time $t_{R}$ shown in Fig.\ref{fig_spatial}. Notice that in the case where the eigen-energies are close together and the band width becomes wider, the PIF 
correction is only a constant pre-factor and the focalization quality by TRM increases.

\section{Comparison with other results}

In a more general point of view, we may develop an alternative derivation of
the PIF formula based on the fundamental optical theorem \cite{Torres}. In
words, it states that the continuous density of states at a given site $x_{0}$
is built upon the escape rates through the boundaries of the system. If we
chose a semi-infinite chain, there is no escape to the left of site $x_{0}$
while an excitation can escape to the right of site $x_{S}$ with a rate
$\operatorname{Im}\Sigma_{x_{S}}/\hbar$. In equations:%
\begin{equation}
\operatorname{Im}G_{x_{0},x_{0}}=G_{x_{0},x_{S}}\operatorname{Im}\Sigma
_{x_{S}}G_{x_{S},x_{0}}^{\ast}.
\end{equation}
Multiplying both sides by a pulse $s(x_{0},\varepsilon)$ and the
$\mathrm{i}\hbar$ factor one finds%
\begin{equation}
\mathrm{i}\hbar\left(  G_{x_{0},x_{0}}-G_{x_{0},x_{0}}^{\ast}\right)
s(x_{0},\varepsilon)=2\mathrm{i}G_{x_{0},x_{S}}\Gamma_{x_{S}}(-\mathrm{i}%
\hbar)G_{x_{S},x_{0}}^{\ast}s(x_{0},\varepsilon).
\end{equation}
If we identify the outgoing wave originated on the excitation at the focal
point as%
\begin{equation}
\psi(x_{0},\varepsilon)=\mathrm{i}\hbar G_{x_{0},x_{0}}s(x_{0},\varepsilon),
\end{equation}
we re-obtain the PIF formula by identifying the second term in the left hand
side as the incoming wave function that produced the pulse in the focal point.
The complete evolution, which is equal to the time reversed one, results%
\begin{subequations}
\begin{align}
\psi(x_{0},\varepsilon)+\psi^{\ast}(x_{0},\varepsilon)  &  =\psi
_{\mathrm{rev}}(x_{0},\varepsilon)\\
&  =\mathrm{i}\hbar G_{x_{0},x_{S}}\underset{{\chi}_{\mathrm{PIF}}%
}{\underbrace{\left[  \frac{2}{\hbar}\Gamma_{x_{S}}\psi^{\ast}(x_{S}%
,\varepsilon)\right]  }}.
\end{align}
The term in square brackets is the excitation that must be injected at $x_{S}$
to produce $\psi_{\mathrm{rev}}$. Notice also that this prescription does not
depend on $x_{0}.$

It is interesting to show some similarities between PIF and the focusing
procedures (STIF and miSTIF) described in Ref.\cite{Vignon}. We will consider
a simple example with a single transducer at $x_{S}$ and a focusing point
located in $x_{0}.$ The goal of STIF is to inject some signal at $x_{S}$ that,
at a later time, is able to form a spatio-temporal delta function in $x_{0}$.
The strategy of STIF is to compute the inverse of the propagator
$G(x_{S},x_{0};\varepsilon)\equiv G_{x_{S},x_{0}}$. The injection proposed by
the STIF could be interpreted as that which achieves the time reversal of an
hypothetical pulse emitted by a source at $x_{0},$ i.e. $s(x,t)=\delta
_{x,x_{0}}\delta(t)$, resulting%
\end{subequations}
\begin{equation}
\chi_{\mathrm{STIF}}(x_{S},\varepsilon)=\tfrac{1}{\mathrm{i}\hbar}%
G_{x_{0},x_{S}}^{-1}\tfrac{\hbar}{2V}s(x_{0},\varepsilon)=\frac{1}%
{\mathrm{i}\hbar}\left[  G_{x_{S},x_{0}}^{\ast}G_{x_{0},x_{S}}\right]
^{-1}G_{x_{S},x_{0}}^{\ast}\tfrac{\hbar}{2V}s(x_{0},\varepsilon).
\end{equation}
Since the time reversal mirror injection is given by%
\begin{equation}
\chi_{\mathrm{TRM}}(x_{S},\varepsilon)=\tfrac{2V}{\hbar}\psi^{\ast}%
(x_{S},\varepsilon),
\end{equation}
and%
\begin{equation}
\psi^{\ast}(x_{S},\varepsilon)=-\mathrm{i}\hbar G_{x_{S},x_{0}}^{\ast}%
s(x_{0},\varepsilon),
\end{equation}
we can rewrite the STIF as a filter%
\begin{equation}
\chi_{\mathrm{STIF}}(x_{S},\varepsilon)=\tfrac{1}{4V^{2}}\left[
G_{x_{S},x_{0}}^{\ast}G_{x_{0},x_{S}}\right]  ^{-1}\chi_{\mathrm{TRM}}%
(x_{S},\varepsilon). \label{eq_STIF}%
\end{equation}
Notice that this filter requires the determination of the non-local Green's
function connecting the transducer at $x_{S}$ and the focal point $x_{0}$.
Therefore, the miSTIF was developed as a procedure that replaces it in terms
of the local Green's function at the transducer site. Such replacement is
enabled by the optical theorem%
\begin{equation}
\operatorname{Im}G_{x_{S},x_{S}}=G_{x_{S},x_{0}}\operatorname{Im}\Sigma
_{x_{0}}^{-}G_{x_{0},x_{S}}^{\ast}+G_{x_{S},x_{S}}\operatorname{Im}%
\Sigma_{x_{S}}^{+}G_{x_{S},x_{S}}^{\ast},
\end{equation}
applied to an infinite chain where $x_{S}>x_{0}$. Here, $\Sigma_{x_{0}}^{-}$
($\Sigma_{x_{S}}^{+}$) is the self-energy correction that accounts for the
free propagation at the left (right) of the focal point (transducer) whose
imaginary part is $\Gamma_{x_{0}}^{-}$ ($\Gamma_{x_{S}}^{+}$). Since%
\begin{equation}
\operatorname{Im}G_{x_{S},x_{S}}\equiv G_{x_{S},x_{S}}\operatorname{Im}%
[\Sigma_{x_{S}}^{+}+\Sigma_{x_{S}}^{-}]G_{x_{S},x_{S}}^{\ast},
\end{equation}
we may use that, in a homogeneous system, the escape at both sides has equal
group velocities%
\begin{equation}
\operatorname{Im}[\Sigma_{x_{S}}^{+}+\Sigma_{x_{S}}^{-}]=-2\Gamma_{x_{S}}.
\end{equation}
In consequence, under these restricted condition, the optical theorem leads to%
\begin{equation}
G_{x_{S},x_{0}}^{\ast}G_{x_{0},x_{S}}=-\frac{1}{2\Gamma_{x_{0}}}%
\operatorname{Im}G_{x_{S},x_{S}},
\end{equation}
from which the miSTIF prescription can be obtained as a broad-band
approximation to Eq.(\ref{eq_STIF}) in which $\Gamma_{x_{0}}(\varepsilon
)=\Gamma_{x_{S}}(\varepsilon)\sim\Gamma=V$,%
\begin{equation}
\chi_{\mathrm{miSTIF}}(x_{S},\varepsilon)=\tfrac{1}{2V}\left[
-\operatorname{Im}G_{x_{S},x_{S}}\right]  ^{-1}\chi_{\mathrm{TRM}}%
(x_{S},\varepsilon), \label{eq_miSTIF}%
\end{equation}
i.e. the minimally invasive spatio-temporal inverse filter.

A comparison between miSTIF and PIF shows that both coincide since the
self-energy correction to the left and right sides of the chain are the same
and%
\begin{subequations}
\begin{align}
G_{x_{S},x_{S}}  &  =\frac{1}{\varepsilon-2\Delta(\varepsilon)+\mathrm{i}%
2\Gamma(\varepsilon)}\\
&  =-\frac{\mathrm{i}}{2\Gamma(\varepsilon)},
\end{align}
and hence, in that particular case%
\end{subequations}
\begin{equation}
\left[  \operatorname{Im}G_{x_{S},x_{S}}\right]  ^{-1}=-\operatorname{Im}%
\left[  G_{x_{S},x_{S}}^{-1}\right]  .
\end{equation}
and we see that both injection prescriptions, the PIF and miSTIF, are proportional.

In turn, since PIF is an exact procedure, it allows to test the efficiency of
miSTIF in focusing the excitation to achieve a spatio-temporal delta function.
This actually results as the back-propagation of the "original" pulse
excitation. As such, in our one dimensional homogenous case, the amplitude at
the focal point is not a delta function in the time domain but $J_{0}%
(2Vt/\hbar)$, a Bessel function of first kind. In the acoustic case, where the
wave function normalization is not required, this can becomes an approximation
to the delta function as $V\rightarrow\infty$.

\section{Conclusions}

We presented the generalization of the PIF focusing method in the internal
source condition. Although we considered the Schr\"{o}dinger equation, since this
formulation is based in general properties of the propagators, as before
\cite{Calvo}, the results remain valid to describe sound propagation. Contrary
to what happens when the source is external, the prescription one has to use
in this case does not involve internal details of the cavity but only simpler
information about the propagation in the outer region. This simplified filter
applies to two physically relevant situations: 1) When the excitation is
actually originated in the interior of the cavity by an external source. In
this case, one obtains a perfect reversal of the wave function for all the
times after the source have been turned-off. 2) Whenever one uses an external
source, but one can separate the incoming from the outgoing waves. In that
case, one can use the recording of the outgoing wave to perfectly reverse the
whole excursion of the excitation through the cavity. In principle, one can
achieve this condition when the boundaries are placed far enough from the
reverberant region. An interesting consequence of our result is the
affirmation that for close cavities in the acoustic case, TRM should produce
perfect time reversal, up to a constant factor, provided that the external
propagation is free of collisions.

We confirm the validity of this expression both analytically and numerically
in a multichannel system and compared with its TRM counterpart. Results on the
reversion quality show that PIF correction constitute a notable improvement
over TRM prescription in cases where the energy dependence has a non-linear
dispersion relation, as typical excitations described by the Schr\"{o}dinger
equation, or when the escape velocity presents structures due to the presence
of many simultaneously propagating channels or when collisions outside the
cavity have a relevant contribution.

\end{document}